\documentclass[aps,prb,reprint,superscriptaddress,amsmath]{revtex4-1}
\usepackage{hyperref}
\usepackage[dvips]{graphicx}

\usepackage{amsmath,amssymb,amsbsy}

\newcommand{\txt}[1]{\mathrm{#1}}

\newcommand{\EC}{E_\txt{c}}

\newcommand{\Ec}{\EC}
\newcommand{\kB}{k_\txt{B}}

\newcommand{\unit}[1]{\ \mathrm{#1}}

\newcommand{\nS}{n_\mathrm{S}}
\newcommand{\fS}{f_\mathrm{S}}
\newcommand{\fN}{f_\mathrm{N}}

\newcommand{\Qdot}{\dot{Q}}
\newcommand{\RT}{R_\mathrm{T}}

\newcommand{\Tenv}{T_\txt{env}}

\renewcommand{\Re}{\mathrm{Re}}

\newcommand{\TS}{T_\mathrm{S}}
\newcommand{\TN}{T_\mathrm{N}}

\newcommand{\be}{\begin{equation}} \newcommand{\ee}{\end{equation}}
\newcommand{\ba}{\begin{eqnarray}} \newcommand{\ea}{\end{eqnarray}}

\newcommand{\ie}{i.\,e.}

\newcommand{\parz}[2]{#1 \left| \right| #2}

\begin{document}

\title{Environmentally Activated Tunneling Events in a Hybrid Single-Electron Box}
\author{O.-P. Saira}
\affiliation{Low Temperature Laboratory, Aalto University, P.O. Box 15100, FI-00076 AALTO, Finland}
\author{M. M\"ott\"onen}
\affiliation{Low Temperature Laboratory, Aalto University, P.O. Box 15100, FI-00076 AALTO, Finland}
\affiliation{Department of Applied Physics/COMP,  Aalto University, P.O.~Box 14100, FI-00076 AALTO, Finland}
\author{V. F. Maisi}
\affiliation{Centre for Metrology and Accreditation (MIKES), P.O. Box 9, 02151 Espoo, Finland}
\author{J. P. Pekola}
\affiliation{Low Temperature Laboratory, Aalto University, P.O. Box 15100, FI-00076 AALTO, Finland}

\begin{abstract}
We have measured individual tunneling events and Coulomb step shapes in single-electron boxes with opaque superconductor-normal metal tunnel junctions. We observe anomalous broadening of the Coulomb step with decreasing temperature in a manner that is consistent with activation of first-order tunneling events by an external dissipative electromagnetic environment. We demonstrate that the rates for energetically unfavourable tunneling events saturate to finite values at low temperatures, and that the saturation level can be suppressed by more than an order of magnitude by a capacitive shunt near the device. The findings are important in assessing the performance limits of any single-electronic device. In particular, master equation based simulations show that the electromagnetic environment realized in the capacitively shunted devices allows for a metrologically accurate charge pump based on hybrid tunnnel junctions.
\end{abstract}
\maketitle

\section{Introduction}
\label{sec:intro}

Various kinds of electron pumps and turnstiles based on single electronics are under active investigation at the moment in order to redefine the unit {\it ampere} and to provide the current source for the quantum metrological triangle experiment~\cite{FeltinPiquemal_PumpEPJ}. The performance requirements in metrology are demanding: To improve the existing uncertainties of fundamental constants, elementary charges have to be transferred at a frequency of the order of 1~GHz with an error rate less than $10^{-7}$. With present-day thin-film technology, it is straightforward to realize electron pumps where the error rates due to thermally activated one-electron processes are below the metrological bound at sub-Kelvin temperatures. Residual pumping errors in normal metal pumps are generally attributed to co-tunneling~\cite{Lotkhov_R3, *Jensen_Accuracy} and environmental activation (EA)~\cite{Covington_PAT, *Keller_WellCharacterized}, although a quantitative theoretical analysis of the experimentally observed error rates is still lacking. A more recently developed candidate is the hybrid turnstile, consisting of a single-electron transistor (SET) with normal metal-insulator-superconductor (NIS) junctions~\cite{Pekola_Pump}. Theoretically, the SINIS turnstile has the potential to reach metrological accuracy~\cite{Averin_PumpTheory}, but the experimental results have been limited by the leakage current present when the turnstile is biased at its operating point~\cite{Lotkhov_RSINIS, *Kemppinen_HighEC}. Leakage in NIS structures is often attributed to a finite density of states (DOS) within the Bardeen-Cooper-Schrieffer (BCS) energy gap, which is modeled with the phenomenological Dynes DOS~\cite{Dynes_QP_Lifetime}. In Ref.~\onlinecite{Pekola_Subgap}, it was shown algebraically that a high temperature environment ($\kB \Tenv \gtrsim \Delta$) can manifest itself as smearing of the DOS in the superconducting electrode of a bare NIS junction, which was experimentally verified. On the contrary, in a Coulomb blockaded system embedded in an environment, the device response depends on one-electron tunneling rates in a way that cannot be expressed in terms of an effective DOS. A tunneling electron can overcome the Coulomb blockade by absorbing energy from the environment, which degrades the performance of the device.

\begin{figure}[t!]
    \begin{center}
        \includegraphics[width=.49\textwidth]{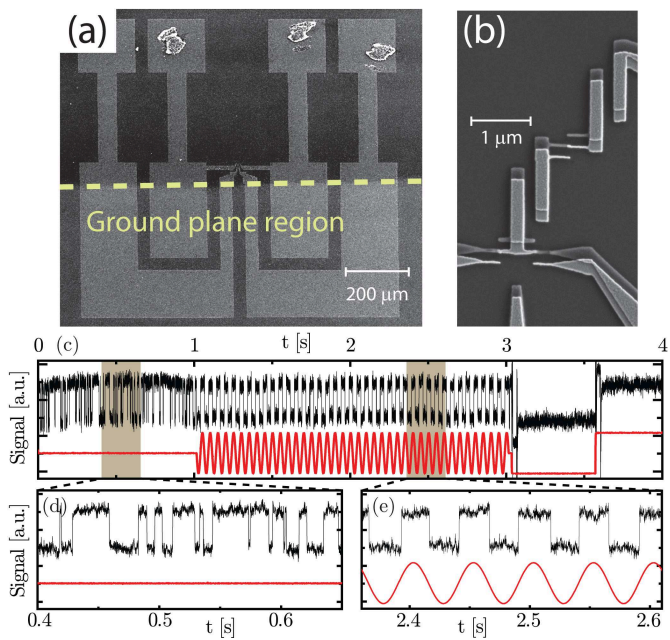}
    \end{center}
   \caption{\label{fig1} (a) Large scale electron micrograph of a sample showing the bonding pads and the electrical leads to the active region in the middle. A copper ground plane isolated from the thin-film metal strucutres by 120~nm aluminium oxide layer is indicated in the bottom half of the picture. (b) Active region of a single-electron box similar to the measured samples. The hybrid single-electron box is shown in the middle, and the electrometer SET below it. The thin-film copper layer appears lighter compared to the aluminium layer. Tips of the gate electrodes can be seen at the bottom and top-right. (c) Four-second real-time electrometer trace (black) recorded from sample G2 demonstrating the ability to accurately control the charge state of the box. The control sequence (red) applied to the box gate is as follows: \mbox{\emph{0-1~s}} Degeneracy point, also detailed in panel (d), showing stochastic switching between the two charge states with nearly equal occupation. \mbox{\emph{1-3~s}} Sinusoidal drive between the charge states at 20~Hz, also detailed in panel (e). \mbox{\emph{3-3.5~s}} and \mbox{\emph{3.5-4~s}} "Hold" mode in the first and second charge state, respectively. The sample stage temperature is 60~mK.}
\end{figure}

In the present work, we study experimentally hybrid tunnel junctions in the EA regime. We demonstrate that EA causes the observed low-temperature saturation of one-electron tunneling rates near the degeneracy point between two charge states, in the energy range that determines the accuracy of the SINIS turnstile when operated with a sinusoidal drive signal. Hybrid tunnel junctions are well-suited for this investigation, as the EA effects are dramatic owing to sharp features in the superconductor DOS, which has been previously exploited in the study of photon-assisted tunneling by microwave irradiation in SETs~\cite{Hergenrother_PAT}. The simplest system where a quantized island charge can be observed is the single-electron box, first studied in Ref.~\onlinecite{Lafarge_Box}. In the present work, we probed the charge state of a galvanically isolated hybrid single-electron box by a capacitively coupled SET electrometer. Owing to the superconducting energy gap, the tunneling rates near degeneracy are sufficiently low so that individual tunneling events can be observed without additional trapping nodes that are required for normal metal devices. We emphasize that this is the same qualitative difference that allows quantized current pumping using the SINIS turnstile with only one Coulomb-blockaded island, whereas the normal metal pumps always have at least two islands. For demonstrating the EA effects, we measured samples with and without a ground plane, \ie, a capacitive shunt in the electrical leads acting as a filter for high-frequency environmental noise. We present data from four samples in total: Samples G1 and G2 have a ground plane as shown in Fig.~\ref{fig1}(a), whereas reference samples R1 and R2 do not have one. Samples G1, G2, and R1 were fabricated simultaneously on the same chip and have identical design for the metal layers. Sample R2 is otherwise similar, but has a SISIS type detector as opposed to the SINIS type found in all the other samples. Figure~\ref{fig1}(b) illustrates the layout of the active region with the single-electron box and the electrometer. 

Two gate electrodes allow for independent tuning of the detector and the box. Whenever the box gate position was changed, a compensation signal was subtracted from the detector gate voltage canceling the direct capacitive coupling to the detector island. For the following discussion, one can consider the compensated box gate voltage to be the only external control signal. Due to the compensation, charge states of the box correspond to fixed detector current levels. The response of the box state to a gate control sequence is demonstrated in Fig.~\ref{fig1}(c)--(e), displaying a real-time measurement of sample G2.

All the tunnel junctions were oxidized strongly, resulting in specific resistances of $3\unit{k\Omega~\mu m^2}$, or about $300\unit{k\Omega}$ for the detector junctions and up to $3.5\unit{M\Omega}$ for the smaller box junction. High resistance is required for the box junction to bring the tunneling rates down into the measurement bandwidth. The detector current was read out in DC-SET configuration using a room-temperature current amplifier, limiting the usable bandwidth to about 5~kHz. 

\section{Theory of environmentally activated tunneling}
\label{sec:model_eat}

Due to small tunnel junctions and the lack of galvanic connection, the charging energy of the box for a single-electron tunneling was at least 4~K in all of the samples. Hence, in all of the experiments performed in dilution refrigerator temperatures, at most two charge states of the box have a non-negligible occupancy. We can safely assume that an equilibrium quasiparticle energy distribution is reached before each tunneling event as the highest tunneling rates considered in the experiment were below 10~kHz. Thus, a complete model for the internal dynamics of the box is a two-state fluctuator, the states of which we will denote by 1 and 2. Neglecting back-action from the detector, the transition rates are symmetric and given by $\Gamma^\pm = \Gamma(\pm E)$, where $E = U_2 - U_1$ is the difference in the electrostatic energy of the two charge states, and $\Gamma(E)$ is the Golden rule tunneling rate~\cite{IngoldNazarov}
\begin{multline}
\label{eq:goldenrule}
\Gamma(E) = \frac{1}{e^2\RT}\int_{-\infty}^\infty  dE_1 \int_{-\infty}^\infty  dE_2 \nS(E_1) \fS(E_1)\\
[1-\fN(E_2)] P(E_1-E_2+E).
\end{multline}
Above, $\nS$ is the DOS for the superconducting electrode, $f_\txt{N(S)}$ is the occupation factor for quasiparticle energy levels in the N(S) electrode, and $P(E')$ gives the probability for emitting energy $E'$ to the environment. In our model, we use the pure BCS density of states $\nS(E) = |\Re \frac{E}{\sqrt{E^2 - \Delta^2}}|$ with $\Delta = 200\unit{\mu e V}$ for the superconducting gap parameter of aluminium. As we will show in Appendix~\ref{sec:qp_thermalization}, the quasiparticles in the N electrode are thermalized by electron-phonon coupling, and in the S electrode through the large overlap with the copper shadow. Hence, the occupation factors are taken to be Fermi functions at sample stage temperature $T_0$, \ie, $f_{N,S}(E) = \left[1 + \exp(E/\kB T_0)\right]^{-1}$. The probability for a two-electron process to change the box charge state is low, as for all those processes the energy cost due to charging energy and Cooper pair breaking is the same or greater, but the rate prefactor is much smaller than for one-electron tunneling.

\section{Model of the electrical environment}
\label{sec:model_zt}

Electromagnetic environment appears in the $P(E)$ theory through the expression $\Re\left[Z_t(\omega)\right]$, where $Z_t(\omega)$ is the total impedance of the environment including the parallel junction capacitance, as seen from the two electrodes of the junction. We model the external environment as a parallel $RC$-element with the resistor at liquid Helium temperature in the same manner as in Refs.~\onlinecite{Pekola_Subgap,Hergenrother_PAT}. Assume that the box electrodes have a mutual capacitance of $C_\txt{j}$, the main contribution of which is from the junction area, and that the two electrodes are coupled to the environment through capacitances $C_1$ and $C_2$, respectively. Then, the impedance $Z_t(\omega)$ can be evaluated as
\begin{equation}
Z_t(\omega) = \parz{\frac{1}{i \omega C_\txt{j}}}{\left(\frac{1}{i\omega C_0} + \parz{R}{\frac{1}{i \omega C}}\right)},
\end{equation}
where $C_0 = \left(1/C_1 + 1/C_2\right)^{-1}$ is the series capacitance of the coupling capacitances $C_1$ and $C_2$, and $\parz{A}{B} := \left(1/A + 1/B\right)^{-1}$ is used to denote parallel impedance. Under the assumption $C_\txt{j} \ll C$, we have the algebraic result
\begin{equation}
\Re\left[Z_t(\omega)\right] = \Re\left[\frac{R/\kappa}{1 + i\omega (R/\kappa) (\kappa C)}\right],\label{eq:Zt_kappa}
\end{equation}
where $\kappa = 1 + C_\txt{j}/C_0$. We refer to $\kappa$ as the decoupling factor, as $\kappa = 1$ corresponds to the case where a bare junction is directly connected to the environment. In the hybrid single-electron box samples, a typical box junction has $C_\txt{j} = 0.2\unit{fF}$ based on the observed charging energies of $4-5\unit{K}$, whereas the couplings are of the order $C_{1,2} = 0.02\unit{fF}$ based on the gate modulation periods and geometric modeling. Hence, one can expect $\kappa$ of the order of 20. It is hard to give an accurate estimate of $\kappa$ based on the conductor geometry as the environmental noise can couple to the box from all of the gate and bias lines, but on the other hand the common-mode component of the noise does not affect tunneling through the junction. We do not try to determine an explicit value for $\kappa$, but instead use $R_\txt{eff} := R/\kappa$ and $C_\txt{eff} := \kappa C$ directly as fitting parameters. When computing the theoretical predictions, the $P(E)$ function for given environmental parameters is evaluated numerically.

\section{Measurement results}
\label{sec:results}

A basic measurable quantity of the box is the charge state occupation probability that depends on the ratio of forward and backward tunneling rates, and assumes the form $p = \Gamma^-/(\Gamma^+ + \Gamma^-)$ for state 1. For an NIN junction under negligible environmental influence, one has the analytic result $\Gamma^\pm = \frac{1}{e \RT} \frac{\pm E}{1 - \exp(\mp E / \kB T_0)}$, giving a Fermi form for the occupation as a function of the energy difference $E$ of the charge states, $p_\txt{NIN} = \left[1 + \exp(E/\kB T_0)\right]^{-1}$. Occupation factors can be determined from the Coulomb staircase, \ie, the detector DC current as a function of the control gate, by normalizing the current levels of consecutive plateaus to 0 and 1, respectively. Because at most two states are active, detector non-linearity with respect to box charge does not affect the results. The occupation can be determined whenever the detector has sufficient sensitivity to the box charge, even when the tunneling rates are beyond the measurement bandwidth. If the tunneling rates are low enough so that the detector can separate the charge states, we calculate the occupation from the weight of histogram peaks which gives improved immunity to low-frequency drifts in the detector signal. 

\begin{figure}[ht!]
    \begin{center}
        \includegraphics[width=.49\textwidth]{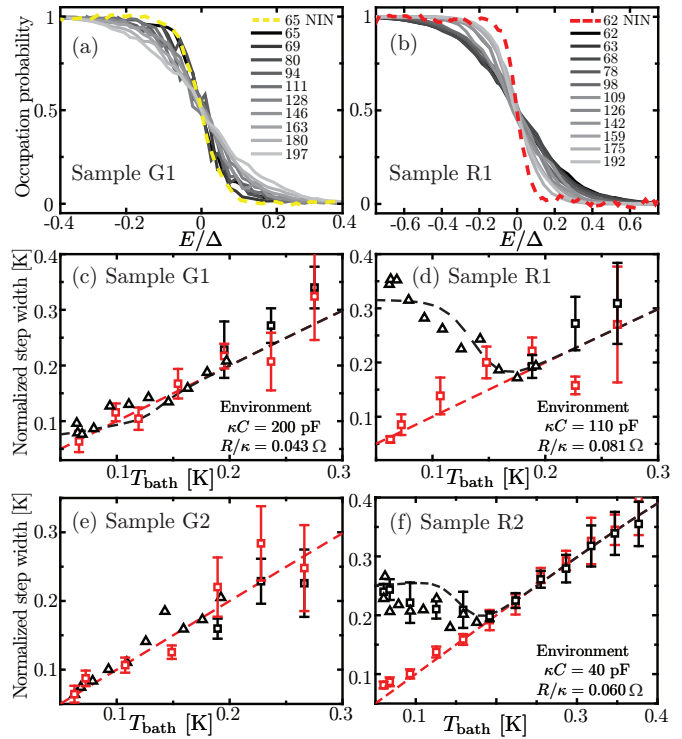}
    \end{center}
   \caption{\label{fig2} (a), (b) Grayscale lines: Coulomb staircase in the NIS state, \ie, charge state occupation probability near the degeneracy point as a function of the energy difference for samples G1 and R1 at temperature points in the range 60--200~mK corresponding to the step width data presented in Fig.~\ref{fig1}(a), (b). The curves shown here are obtained by analyzing real-time electrometer traces where the two states are separable. For sample G1 with the ground plane, the step sharpens with reduced temperature, whereas for the reference sample R1 the opposite is observed. Colored dashed line: Staircase in the NIN state at the base temperature. (c)-(f) Step widths as a function of temperature in NIS state (black symbols) and NIN state (red symbols). Black triangles correspond to step widths determined using real-time traces. In NIN state (red squares), and in NIS state at elevated temperatures (black squares), the step widths have been determined from electrometer DC current averaged over the charge state fluctuations. Dashed lines show the predictions of $P(E)$ theory for NIS state (black) and NIN state (red).}
\end{figure}

\begin{figure}[t!]
    \begin{center}
        \includegraphics[width=.49\textwidth]{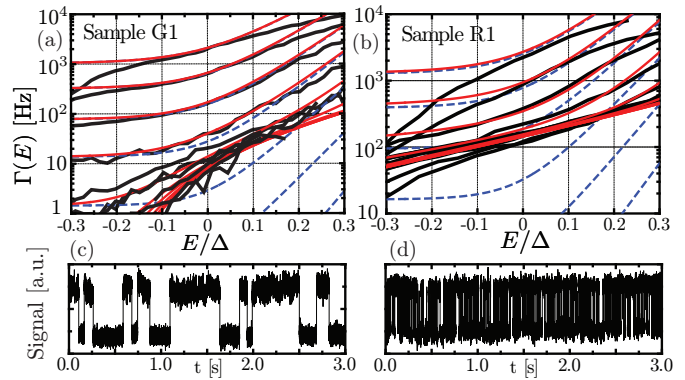}
    \end{center}
   \caption{\label{fig3} (a), (b) Tunneling rates as a function of energy difference for samples G1 and R1, respectively, at temperatures in the range 60--200~mK corresponding to the NIS state data presented in Fig.~\ref{fig2}(a), (b). Black curves are experimental data whereas red curves show the theoretical predictions using the same parameter values as in Fig.~\ref{fig2}. For reference, we plot orthodox theory rates without environment as dashed blue lines. The experimental rates are most reliable near the degeneracy point. Counting errors due to missed events increase as one of the tunneling rates grows beyond the detector bandwidth, explaining the discrepancy between experiment and theory for sample R1 at high temperatures. (c), (d) Individual traces at the base temperature near the degeneracy point for samples G1 and R1, respectively, demonstrating the reduction of tunneling rates by capacitive shunting.}
\end{figure}

We determine the width of a step shape obtained from experiment or numerical calculation by fitting a Fermi function to it. Numerical results indicate that for the environmental parameters considered in the present work, the Coulomb steps at high temperatures in both NIN and NIS states have the Fermi form given above, with the width corresponding to the common temperature of the bath and the quasiparticles in both electrodes. In NIN state, the linear relation between step width and temperature is valid down to our base temperature of 60~mK. However, in NIS state the step width behaves anomalously at temperatures below 200~mK with strong dependence on the environmental parameters. Experimentally, the hybrid box can be brought into NIN state by an external magnetic field.  In Fig.~\ref{fig2} we show the data from all four samples measured in the temperature range 60--300~mK (60--400~mK for sample R2) in both NIS and NIN states. The NIN data in panels (c)--(f) for all samples displays linear behavior down to the base temperature, indicating that the box quasiparticle temperatures were not elevated due to external noise heating. The energy difference $E$ depends linearly on the control gate voltage, with the conversion factor determined by the coupling capacitances and the box charging energy. We determined the conversion factor from gate voltage offset into energy by requiring that the step widths in the linear regime correspond to energy $\kB T_0$. For samples G1 and R1, we display theoretical predictions using environmental parameters that yield best fit to the data presented in both Figs.~\ref{fig2} and \ref{fig3}. For sample R2, provisional parameters reproducing the observed step widths are used. The data for sample G2 can be reproduced by parameters that are similar to those of sample G1. We note that if at least part of the environment noise is capacitively coupled, adding shunting capacitance over the sample will also decrease the effective source resistance.

In the NIS data for the unshielded samples R1 and R2, EA effects can be observed at temperatures below 200~mK as broadening of the step with decreasing temperature. At low temperatures, the tunneling rates for small energy differences are set by EA, and consequently the step width is determined by the environmental parameters. When the temperature is sufficiently high, thermal activation (TA) due to broadening of the quasiparticle energy distribution begins to govern the tunneling rates, and a linear relation between step width and temperature is restored. The EA-limited step width can correspond to a higher temperature than where the transition to TA regime occurs, which results in a width minimum at finite temperature. The measured step width for ground plane samples G1 and G2 is nearly linear for the whole experimental temperature range which is reflected in smaller $R_\txt{eff}$ and larger $C_\txt{eff}$ for the environment.

We determined the tunneling rates by stepping the control gate voltage across the degeneracy point, identified as the midpoint of a Coulomb staircase step. At each gate position, a three-second trace of the output of the detector current amplifier was recorded at 100~kHz sampling rate followed by digital low-pass filtering with a sharp cut-off at 5~kHz. The whole sweep was discarded if an abrupt change in the detector or box operating point due to shifts in background charges was observed. The tunneling events were identified using threshold detection, with the transition thresholds set based on the histogram computed for a trace. The tunneling rates are estimated as $\Gamma^{\pm} = \bar{\tau}_{1(2)}^{-1}$, where $\bar{\tau}_{1(2)}$ is the mean observed lifetime of state 1(2). Due to the possibility of missed events, the rate estimates obtained from a trace are reliable only if both rates satisfy $1/t_0 \ll \Gamma^{\pm} \ll B_\txt{det}$, where $t_0$ is the length of the trace and $B_\txt{det}$ is the detector bandwidth.

In Fig.~\ref{fig3}(a)--(b), we present the measured tunneling rates at different bath temperatures in the range 60--200~mK for the shielded sample G1 and the reference sample R1. We also show the $P(E)$ theory predictions using the same environment parameters as in Fig.~\ref{fig2}. For comparison, the tunneling rates without EA are also shown. The tunneling resistance $\RT$ of the box junction, an otherwise unaccessible parameter, was determined from the data at 180~mK where tunneling rates depend weakly on the environmental parameters, but are still within the detector bandwidth. The values obtained this way are $3.5\unit{M\Omega}$ for sample G1 and $2.0\unit{M\Omega}$ for R1. In the tunneling rate data, the agreement is good for the two samples in the $E$ range where both rates $\Gamma^\pm = \Gamma(\pm E)$ are below 1~kHz. Above that, the experimentally determined rates are smaller than the real values due to missed events. For both samples, the floor where low-temperature rates saturate to is set by the environment. In sample R1, the tunneling rates recorded at 125 and 140~mK are lower than the base temperature rates, which we interpret as an artifact due to reduced sensitivity of the detector SET at higher temperatures. As illustrated in Fig.~\ref{fig3}(c)--(d), in the ground plane sample G1 the base temperature tunneling rate at the degeneracy point is an order of magnitude smaller than in the unshielded sample R1, even after accounting for the difference in tunneling resistances.

\section{Discussion}
\label{sec:discussion}

To relate the observed tunneling rate data to the pumping performance of the SINIS turnstile, we performed master equation based simulations of electron transfer during the pumping cycle. For the simulations, we assumed a high-quality turnstile with realistic parameters $\RT = 0.5\unit{M\Omega}$ per junction, $\Ec/\kB = 5\unit{K}$, and a quasiparticle temperature of 50~mK for all electrodes. Gate-dependent first-order tunneling rates were calculated according to Eq.~(\ref{eq:goldenrule}) using the environment parameters that we extracted for the shielded sample G1. At 20~MHz pumping frequency, the simulations predict an EA-limited total error rate of $1.7\times10^{-6}$ for a sinusoidal gate drive, and $1.8\times10^{-7}$ for a square wave drive. Cancellation of pumping errors of different signs is not included in these figures, and hence the relative deviation of the average pumped current from the ideal value $I = e f$ is smaller. It  remains an open technological challenge to realize a similar or better environment for an SET with galvanic connections to biasing leads.  

In summary, we have measured single-electron tunneling rates in an NIS single-electron box near the degeneracy point in two qualitatively different electromagnetic environments. The agreement to a model of first order tunneling activated by absorption of photons from a warm environment is excellent. Accounting for the enhanced decoupling in the present experiment due to lack of galvanic connections, the parameter values we obtained for the external environment are comparable with those presented in Ref.~\onlinecite{Pekola_Subgap}. The results indicate that environmental fluctuations deteriorate the performance of single-electronic devices due to enhancement of tunneling rates in the energetically unfavourable direction. The measurements on the sample with an on-chip capacitive shunt indicate that the environment-induced error rate of the SINIS turnstile can be suppressed to the metrological level with an appropriate filtering of high-frequency noise. We emphasize that the EAT model explains the observed tunneling rates without any sub-gap states in the superconductor. 

\begin{acknowledgments}
We thank M. Meschke and S. Kafanov for helpful discussions. The work has been supported partially by the Academy of Finland, V\"ais\"al\"a Foundation, Emil Aaltonen Foundation, and the Finnish National Graduate School in Nanoscience. The research conducted within the EU project SCOPE received funding from the European Community's Seventh Framework Programme under Grant Agreement No. 218783.
\vspace{1cm}
\end{acknowledgments}

\appendix
\section{Detector back-action}
\label{sec:backaction}

Typical value for the average detector current used in the real-time counting experiment was 0.5~nA. Low detector current has the benefit of reduced heating and back-action at the cost of signal-to-noise ratio. To assess experimentally the effect of detector back-action, we measured the tunneling rates at base temperature at different operation points of the detector. In the parameter range where tunneling events could still be reliably detected, we observed that the tunneling rate at the degeneracy point for the shielded sample G2 was modulated by factor three. For the unshielded sample R1, we found no change within experimental error. We were able to reproduce these results qualitatively using a model similar to Ref.~\onlinecite{Turek_SETBackAction}, where the effective position of the box gate fluctuates due to changes in the charge state of the detector island. Detector back-action is not included in the theoretical curves of Figs.~2 and 3, which may affect the values of fitted environment parameters for the shielded sample, but does not change the conclusions of the paper about the role of the external environment.

\section{Quasiparticle thermalization in the box electrodes}
\label{sec:qp_thermalization}

Here, we address the issue of quasiparticle heating and thermalization in the electrodes of the hybrid single-electron box. The heat associated with quasiparticle tunneling governed by Eq.~(\ref{eq:goldenrule}) can be calculated as a function of the charging energy difference $E$ as
\begin{widetext}
\begin{eqnarray}
\dot{Q}_\txt{S}(E) & = -\frac{1}{e^2\RT}\int_{-\infty}^\infty dE_1\, E_1 \int_{-\infty}^\infty dE_2\, \nS(E_1) \fS(E_1)[1-\fN(E_2)] P(E_1-E_2+E), \label{eq:QdotS}\\
\dot{Q}_\txt{N}(E) & = \frac{1}{e^2\RT}\int_{-\infty}^\infty  dE_1 \int_{-\infty}^\infty dE_2\, E_2\, \nS(E_1) \fS(E_1)[1-\fN(E_2)] P(E_1-E_2+E).\label{eq:QdotN}
\end{eqnarray}
\end{widetext}
for the S and N electrodes, respectively. The average energy deposited per individual tunneling event can be evaluated as $\Qdot_\txt{N,S}(E) / \Gamma(E)$. For the single-electron box with two active charge states, the average heat load is given by
\begin{equation}
\left<\Qdot_\txt{N,S}\right> = \frac{\Qdot_\txt{N,S}(E)/\Gamma(E) + \Qdot_\txt{N,S}(-E)/\Gamma(-E)}{1/\Gamma(E) + 1/\Gamma(-E)}.\label{eq:QdotAvg}
\end{equation}
Numerical calculations show that the maximum heating occurs at the degeneracy point, \ie, $E = 0$. Using the sample and environment parameters extracted from the tunneling rate data, the heat loads in samples R1 and G1 calculated from Eqs.~(\ref{eq:QdotS})--(\ref{eq:QdotAvg}) are presented in Tbl.~\ref{tbl1}.

In the N electrode, hot quasiparticles can thermalize through electron-phonon coupling. The electron-phonon heat flow is given by~\cite{Giazotto_RMP}
\begin{equation}
\Qdot_\txt{e-ph} = \Sigma \mathcal{V} (T_\txt{N}^5 - T_\txt{N,ph}^5),
\end{equation}
where $\Sigma$ is the material-dependent electron phonon coupling constant, $\mathcal{V}$ is the island volume, and $T_\txt{N,ph}$ is the temperature of the island phonons, assumed to equal the sample stage temperature $T_0$. We use the literature value~\cite{Giazotto_RMP} for copper $\Sigma = 2\times10^9 \unit{Wm^{-3}K^5}$, and $\mathcal{V} = 0.3 \times 1.5 \times 0.04\unit{\mu m^3}$. The island phonons are assumed to be well thermalized to the sample stage. Using the heat load values presented in~Tbl.~\ref{tbl1} at the base temperature $T_0 = 60\unit{mK}$, we find that the N electrode quasiparticle temperature is elevated at most $0.02\unit{mK}$.
 
In the S electrode, the electron-phonon heat flow is exponentially suppressed below the critical temperature~\cite{Timofeev_BCS_EP}. Hence, the dominant heat conduction channel is through the $0.3\times1.3 \unit{\mu m^2}$ overlap with the copper shadow. Using the specific resistance value $3\unit{k\Omega~\mu m^2}$ obtained for the SET junctions, we get $\RT = 8\unit{k\Omega}$ for the overlap junction. We obtain a lower bound for the heat flow by considering only first-order quasiparticle tunneling. In order to determine the quasiparticle saturation temperature, henceforth denoted by $\TS^0$, we then evaluate the heat flow though the junction from Eq.~(\ref{eq:QdotS}) as a function of $\TS$ with $\TN = 60\unit{mK}$. Environmental fluctuations can be neglected due to large junction capacitance, \ie, we use $P(E) = \delta(E)$. For the S electrode heat loads presented in Tbl.~\ref{tbl1}, we find that $\TS^0 = 116\unit{mK}$ for sample R1, and $100\unit{mK}$ for sample G1. However, the tunneling rates in the parameter range accessed in the experiment have a very weak dependence on $\TS$. The theoretical curves presented in the article are calculated with $\TS = T_0$, and there is a negligible difference between them and curves calculated with $\TS = \TS^0$.

\begin{table}
\caption{\label{tbl1} Calculated heat loads to the box electrodes from quasiparticle tunneling in various configurations at 60~mK.}
\begin{ruledtabular}
\begin{tabular}{llll}
Sample & NIN state            & NIS state, to N & NIS state, to S\\
R1     & $4.7\times 10^{-20}\unit{W}$ & $1.2 \times 10^{-21}\unit{W}$  & $5.3 \times 10^{-21}\unit{W}$\\
G1     & $1.7\times 10^{-20}\unit{W}$ & $5.5 \times 10^{-24}\unit{W}$ & $2.4 \times 10^{-22}\unit{W}$
\end{tabular}
\end{ruledtabular}
\end{table}

\bibliography{electron_pump}

\end{document}